\documentclass[prd,superscriptaddress,twocolumn,showpacs,nofootinbib]{revtex4}
\usepackage{graphicx}
\usepackage{amsfonts,latexsym}
\usepackage{amsmath,amssymb}
\usepackage{graphicx,color}
\usepackage{multirow}
\usepackage{mathdots}

\newcommand{\be}{\begin{equation}}
\newcommand{\ee}{\end{equation}}
\begin{document}
\input epsf
\title{ Rank-$n$ logarithmic conformal field theory in the BTZ black hole}

\author{Taeyoon Moon}\email{tymoon@sogang.ac.kr}
\affiliation{Center for Quantum Spacetime, Sogang University, Seoul
121-742, Korea}

\author{Yun Soo Myung}\email{ysmyung@inje.ac.kr}
\affiliation{Institute of Basic Sciences and School of Computer
Aided Science, Inje University, Gimhae 621-749, Korea}

\begin{abstract}
We construct the rank-$n$ finite temperature logarithmic conformal
field theory (LCFT) starting from the $n$-coupled scalar field
theory in the BTZ black hole background.  Its zero temperature limit
reduces to  a rank-$n$ LCFT in the AdS$_3$ background whose gravity
dual is  a polycritical  gravity. We compute all two-point functions
of a rank-$n$ finite temperature LCFT.  Using  the retarded Green's
functions on the boundary,  we obtain  quasinormal modes of scalar
 $A_n$ which satisfies the $2n$-th order linearized equation.
Furthermore,  the absorption cross section of $A_n$  indicates a
feature of $\ln^{n-1}[\omega \ell]$-correction to the Klein-Gordon
mode.
\end{abstract}
\maketitle
Critical gravity
based on higher-curvature terms in the AdS$_{\rm
d+1}$-spacetimes~\cite{Li:2008dq,Lu:2011zk,Deser:2011xc,Porrati:2011ku}
  has regarded as  a toy model for quantum
gravity.  At the critical point of avoiding the ghosts, a degeneracy
takes place and massive gravitons coincide with massless gravitons.
All massive gravitons are replaced by  an equal amount of
logarithmic modes  at the critical point, leading to the critical
gravity (log-gravity). According to the dictionary of AdS-LCFT
correspondence, one finds that a rank-2 logarithmic conformal field
theory (LCFT) is dual to a critical
gravity~\cite{Grumiller:2008qz,Myung:2008dm,Maloney:2009ck}.

However, one has to face with the non-unitarity issue of the
log-gravity theory because it contains higher-derivative terms. In
order to resolve this issue, a polycritical gravity with
$2n(n>2)$-derivative was introduced to provide multiple critical
points~\cite{Nutma:2012ss} whose CFT dual seems to be a rank-$n$
LCFT. A consistent unitary truncation of polycritical gravity was
performed  at the linearized level for odd $n$, but not for even
$n$~\cite{Kleinschmidt:2012rs}.

On the other hand, a $n$-coupled scalar field model in the
 AdS$_{\rm d+1}$-spacetimes has been proposed as a toy model for a
 gravitational dual to a rank-$n$ LCFT~\cite{Bergshoeff:2012sc}.
 Introducing $n-1$ auxiliary scalar fields, this model could be
 rewritten as a two-derivative theory. The critical point is obtained when  all masse of the $n$-scalar
 fields degenerate. The $n-1$ higher-order logarithmic modes appear
 as  logarithmic partners of the Klein-Gordon scalar $A_1$. This model is compared to the
 $2n$-derivative Lee-Wick model where the odd $n$ and the even $n$
 cases feature qualitative difference~\cite{Carone:2008iw}.

In this Report, we will investigate a $n$-coupled  scalar field
model on the BTZ black hole background~\cite{Banados:1992wn}. Our
purpose is twofold. One is to
 recognize the difference in the AdS-LCFT correspondence between a $n$-coupled
 scalar field model on the AdS$_3$--a rank-$n$ (zero temperature) LCFT and    a $n$-coupled  scalar
field model on the BTZ black hole--a rank-$n$ finite
temperature~LCFT.
 The other is to compute the
quasinormal frequencies of a field $A_n$ satisfying
$(\nabla^2_B-m^2)^n A_n=0$ and the absorption cross section from
scattering $A_n$ off the BTZ black hole by using the retarded
Green's function of rank-$n$ finite temperature LCFT.  Actually,
this computation is a formidable task when using a single scalar
equation \begin{equation} \label{neq}
 (\nabla^2_B-m^2)^n \varphi=0. \end{equation}
    Instead, employing the
$n$-coupled scalar field model to give $(\nabla^2_B-m^2)^n A_n=0$
eventually, one could compute the quasinormal frequencies, and
absorption cross section in the low-temperature and massless limits.

We take a toy scalar model for the  polycritical gravity  with
$n$-coupled scalar fields with degenerate masses
\begin{eqnarray}\label{maina}
S=\int d^{3}x \sqrt{-g}\Big[R-2\Lambda\Big]+S_{\Phi},
\end{eqnarray}
where $S_{\Phi}$ is given by
\begin{eqnarray} \label{sPhi}
S_{\Phi}=-\frac{1}{2}\int d^{3}x \sqrt{-g}\sum_{i,j=1}^{n}\Big[
\alpha_{ij}\partial_{\mu}\Phi_{i}\partial^{\mu}\Phi_{j}
+\beta_{ij}\Phi_i\Phi_j\Big]
\end{eqnarray}
with the $n$-dimensional matrices $\alpha_{ij}$ and $\beta_{ij}$
~\cite{Bergshoeff:2012sc}. Now we introduce a background metric
$\bar{g}_{\mu\nu}$ of the BTZ black hole with the cosmological
constant $\Lambda=-1/\ell^2=-1$ \cite{Banados:1992wn}
\begin{eqnarray}\label{btzm}
&&ds^2_{\rm B}~=~\bar{g}_{\mu\nu}dx^{\mu}dx^{\nu}
~=~-\frac{(r^2-r_+^2)(r^2-r_-^2)}{r^2}dt^2+\nonumber\\
&&\hspace*{1em}\frac{r^2}{(r^2-r_+^2)(r^2-r_-^2)}dr^2
+r^2\left(d\phi+\frac{r_+r_-}{r^2}dt\right)^2.
\end{eqnarray}
Here, the ADM mass $M=r_+^2-r_-^2$, angular momentum $J=2r_+r_-$,
and right/left temperature $T_{R/L}=(r_+\pm r_-)/2\pi$.

 We
consider the perturbation around the background spacetimes
(\ref{btzm}) of  $\Phi_{i}=\bar\Phi_{i}+A_{i}$ with
$\bar\Phi_{i}=0$. Then the linearized equations are given by
\begin{eqnarray}
(\nabla_{B}^2-m^2)A_1=0,~~~(\nabla_{B}^2-m^2)A_p=A_{p-1}\label{eq2}
\end{eqnarray}
with $p=2,\cdots,n$, which lead to the $2n$-th order differential
equation for $A_n$ as
\begin{eqnarray} \label{aa2}
(\nabla_{B}^2-m^2)^nA_n=0.
\end{eqnarray}
It is noted that unlike Eq.(\ref{neq}), Eq.\eqref{aa2} was obtained
from the recursion relation, which means that equation for $A_{p}$
is recursively related to $A_1$.

We are in a position to  compute the bulk-to-boundary propagators so
that a rank-$n$ finite temperature LCFT is formed on the boundary.
The bulk scalar $A_i$ is  represented by  bulk-to-boundary
propagators $K_{ij}$  which relate the bulk solution to the boundary
source fields $A_{i(b)}$. The propagator $K_{ij}$ is given as
\begin{equation}
K_{ij}=\left(
  \begin{array}{ccccc}
    0 & 0 & \cdots & 0 & K_1\\
    0 & 0 & \cdots & K_1 & K_2\\
    \vdots & \iddots & \iddots & \vdots & \vdots\\
    K_1 & K_{2} & \cdots & K_{n-1} & K_{n}\\
  \end{array}
\right) \label{kij}.
\end{equation}
Here $K_{1}$ and $K_{p}$ with $p=2,\cdots,n$ are
\begin{eqnarray}
&&K_{1}=K_{1n}=K_{2~n-1}=\cdots =K_{n1},\label{knn}\\
&&K_{p}=K_{pn}=K_{p+1~n-1}=\cdots =K_{np},\label{knp}
\end{eqnarray}
which satisfy the following relations:
\begin{eqnarray}
(\nabla^2_{\rm B}-m^2)K_1=0,~~~
 (\nabla^2_{\rm B}-m^2)K_p=K_{p-1}.\label{kp}
\end{eqnarray} In this case, $K_{1}$
and $K_p$ correspond to the bulk-to-boundary propagators of the
Klein-Gordon mode $A_1$ and  mode $A_p$, respectively. Importantly,
$K_n$-propagator satisfies
\begin{equation}
(\nabla^2_{\rm B}-m^2)^nK_n=0.
\end{equation}
The solutions $A_{i}(r,u_+,u_-)$ to (\ref{eq2}) are written as
 \begin{eqnarray}\label{AI}
A_{i}= \int du_+^{\prime}du_-^{\prime}\Big[\sum_{j=1}^{n}
K_{ij}(r,u_+,u_-;u_+^{\prime}, u_-^{\prime})A_{j(b)}\Big]
\end{eqnarray}
with $u_\pm=\phi\pm t$.
 It is well-known that in the BTZ black hole background,
$K_1$ can be found to be  the solution to the  Klein-Gordon
equation~\cite{KeskiVakkuri:1998nw}
\begin{eqnarray}
&&K_1(r,u_+,u_-;u_+^{\prime},u_-^{\prime})~=~\nonumber\\
&&\hspace*{-1.1em}\left[\frac{N\pi^2 T_R T_L}{Me^{\pi T_L\delta
u_++\pi T_R\delta u_-}/4r+r\sinh(\pi T_L\delta u_+)\sinh(\pi
T_R\delta
u_-)}\right]^{\triangle}\nonumber\\
&&\equiv
N\Big[f(r,u_+,u_-;u_+^{\prime},u_-^{\prime})\Big]^{\triangle}\label{k1eq},
\end{eqnarray}
where $\delta u_{\pm}=u_{\pm}-u_{\pm}^{\prime}$,
$\triangle(\triangle-2)=m^2$, and $N$ is a normalization constant.
Here the Hawking temperature $T_H$ is defined by $T_{H}=2/(1/T_R
+1/T_L)$. It is worth noting that a general formula for the
bulk-to-boundary propagator $K_p$ leads to
\begin{eqnarray}\label{KP}
&&\hspace*{-1em}K_p=\frac{1}{(p-1)!}\frac{d^{p-1}}{(dm^2)^{p-1}}K_1=
\frac{K_1}{2^{p-1}(p-1)!(\triangle-1)^{p-1}}\nonumber\\
&&\times \Bigg[\ln^{p-1}[f]+\frac{1}{N}\Big\{\sum_{l=1}^{p-1}
{}_{p-1}C_{l}\Big(\frac{\partial^lN}{\partial\triangle^l}\Big)\ln^{p-l-1}[f]
\Big\}\nonumber\\
&&\hspace*{-1.1em}+\frac{1}{(\triangle-1)^{k}N}
\Big\{\sum_{k=1}^{p-2}b_{kp}\sum_{l=0}^{p^{\prime}}
{}_{p^{\prime}}C_{l}\Big(\frac{\partial^lN}{\partial\triangle^l}\Big)
\ln^{p^{\prime}-l}[f] \Big\}\Bigg],
\end{eqnarray}
where $p=2,\cdots,n$, ${}_{n}C_{i}\equiv\frac{n!}{i!(n-i)!}$,
$b_{kp}$ are constant, and $p^{\prime}=p-k-1$. Here we observe the
highest order logarithmic term of ``$\ln^{p-1}[f]K_1$", showing that
the bulk-to-boundary operator is determined as the solution to the
$2p$-th order differential equation $(\nabla^2_B-m^2)^pK_p=0$. Now
we consider an on-shell bilinear action $S_{{\rm eff}}$ on the
boundary:
\begin{equation}\label{seff}
 2 S_{{\rm eff}}=
-\lim_{r_s\to\infty} \int_{s}du_+du_-\sqrt{-\gamma}
[\sum_{i,j=1}^{n}\alpha_{ij}A_i(\hat{n}\cdot\nabla) A_j],
\end{equation}
which leads to  a complicated expression (see arXiv:1211.3679v1)
obtained from inserting Eq.(\ref{AI}) with (\ref{k1eq}) and
(\ref{KP}) into (\ref{seff}). Following the AdS-LCFT correspondence,
we wish to couple the boundary values $A_{i(b)}$ of the fields to
their dual operators ${\cal O}_i$ as $\int du_+du_-\sqrt{-\gamma}
\sum_{i=1}^n A_{i(b)}{\cal O}_i$ for the symmetric-type coupling.
This shows clearly the one-to-one correspondence between $A_{i(b)}$
and ${\cal O}_i$. One can derive the two-point functions for the
dual conformal operators ${\cal O}_{i}$ as follows:
\begin{eqnarray}
&&\hspace*{-1.5em}\Big<{\cal O}_{i}(u_+,u_-){\cal O}_{j-i}(0)\Big>
=\frac{\delta^2S_{\rm eff}}{\delta A_{i(b)}(u_+,u_-)\delta
A_{j-i(b)}(0)}=0,\nonumber\\
\nonumber\\
&&\hspace*{-1.5em}\xi_{kk^{\prime}}\Big<{\cal O}_{k}(u_+,u_-){\cal
O}_{k^{\prime}}(0)\Big> =\frac{\delta^2S_{\rm eff}}{\delta
A_{k(b)}(u_+,u_-)\delta
A_{k^{\prime}(b)}(0)}\nonumber\\
&&\hspace*{-1.2em} = 2^{-\delta_{kk^{\prime}}}\triangle
N\left(\frac{\pi T_L}{\sinh[\pi T_Lu_+]}\right)^{\triangle}
\left(\frac{\pi T_R}{\sinh[\pi T_R u_-]}\right)^{\triangle},
\label{o13}
\end{eqnarray}
where $k^{\prime}=n-k+1$ with $k=1,\cdots,n$ and $i=1,\cdots,n-1$,
$~j=i+1,i+2,\cdots,n.$ For $s=p+q-n-1~>0$ with $p,q=2,\cdots,n$, the
two-point functions are given by
\begin{eqnarray}
\zeta_{pq}\Big<{\cal O}_{p}(u_+,u_-){\cal O}_{q}(0)\Big>
&=&\frac{\delta^2S_{\rm eff}}{\delta A_{p(b)}(u_+,u_-) \delta
A_{q(b)}(0)}\nonumber\\
&&\hspace*{-12em}~=~\triangle N[\tilde{f}(1)]^{\triangle}2^{-\delta_{pq}}\times\nonumber\\
&&\hspace*{-12em}\Bigg\{\frac{1}{2^{s}(s)!(\triangle-1)^{s}}
\Bigg(\ln^s [\tilde{f}(\epsilon)]+\frac{1}{N}\sum_{l=1}^{s}{}_{s}a_l
\ln^{s-l}[\tilde{f}(\epsilon)]\nonumber\\
&&\hspace*{-11em}+\frac{s}{\triangle}
\ln^{s-1}[\tilde{f}(\epsilon)]+\frac{1}{\triangle
N}\sum_{l=1}^{s}(s-l){}_{s}a_l
\ln^{s-l-1}[\tilde{f}(\epsilon)]\nonumber\\
&&\hspace*{-11.5em}+\frac{1}{(\triangle-1)^{k}N}
\Big\{\sum_{k=1}^{s-1}b_{k~s+1}\sum_{l=0}^{s-k} {}_{s-k}a_{l}
\ln^{s-l-k}[\tilde{f}(\epsilon)]+\frac{1}{\triangle}\times\nonumber\\
&&\hspace*{-11.5em}\sum_{k=1}^{s-1}b_{k~s+1}\sum_{l=0}^{s-k}
(s-l-k){}_{s-k}a_{l} \ln^{s-l-k-1}[\tilde{f}(\epsilon)]
\Big\}\Bigg)\Bigg\},\label{opq}
\end{eqnarray}
where $\tilde{f}(\epsilon)$ and ${}_{j}a_{l}$ are given by
\begin{eqnarray}
\tilde{f}(\epsilon)=\epsilon \frac{\pi^2T_RT_L}{\sinh[\pi T_L\delta
u_+]\sinh[\pi T_R\delta u_-]},~{}_{j}a_{l}={}_j
C_{l}\Big(\frac{\partial^{l}N}{\partial\triangle^{l}}\Big).\nonumber
\end{eqnarray}
All correlation functions are zero for $s<0$. We note that this is
one of our main results. In (\ref{o13}) and (\ref{opq}), when
reducing to the AdS$_3$ background \cite{Bergshoeff:2012sc},  the
parameters $\xi_{kk^{\prime}}$ and $\zeta_{pq}$ are determined  to
be
\begin{eqnarray}
\xi_{kk^{\prime}}=2^{-\delta_{kk^{\prime}}}\frac{\triangle
N}{[2(\Delta -1)]^{n}},~~~\zeta_{pq}=\frac{2^{-\delta_{pq}}\triangle
N}{[2(\Delta -1)]^{n+s}}.\nonumber
\end{eqnarray}
At this stage, we wish to point out that if one truncates the theory
with odd rank to be unitary, the only non-zero correlation function
is given by a reduced
 matrix as
 \begin{equation}
  \label{tmatrix}
 <{\cal O}^i{\cal O}^j> \sim \left(
  \begin{array}{cc}
    0& 0 \\
    0 & {\rm CFT} \\
  \end{array}
\right),
 \end{equation}
 which implies that the remaining sector involves a non-trivial
 two-point correlator
\begin{equation}
<{\cal O}^{{\rm log}^{\frac{n-1}{2}}}(x) {\cal O}^{{\rm
log}^{\frac{n-1}{2}}}(0)>
=\frac{[2(\triangle-1)]^n}{|x|^{2\triangle}}.
\end{equation}
 This defines a unitary CFT and thus, the non-unitary issue could be resolved by truncating  a
 rank-$n$ LCFT with $n=3,5,\cdots$.
On the contrary, for the even rank of $n=4,6,\cdots$, it reduces to
a null matrix
\begin{equation}
  \label{tmatrix}
 <{\cal O}^i{\cal O}^j> \sim \left(
  \begin{array}{cc}
    0& 0 \\
    0 & 0 \\
  \end{array}
\right),
 \end{equation}
which contains null states only.

On the other hand, the retarded Green's functions are defined by
\begin{equation}
{\cal D}^{\rm ret}_{jk}(t,\phi;0,0)=i\Theta(t-0)\bar{\cal
D}_{jk}(u_+,u_-),~~j,k=1,2,\cdots,n\nonumber
\end{equation}
where the commutator evaluated in the equilibrium canonical ensemble
are given by
\begin{eqnarray}
\bar{\cal D}_{jk}=\Big<{\cal O}_{j}(u_+ -i\epsilon,-u_-
-i\epsilon){\cal
O}_{k}(0)\Big>~-~\Big<\epsilon\to-\epsilon\Big>.\nonumber
\end{eqnarray}
Making the Fourier transform of  $\bar{\cal D}_{jk}(u_+,u_-)$,
$\bar{\cal D}_{jk}(p_+,p_-)$ takes the form
\begin{eqnarray}\label{Dinp}
\bar{\cal D}_{jk}=\int du_+ du_- e^{i(p_+ u_+-p_- u_-)}\bar{\cal
D}_{jk}(u_+,u_-)
\end{eqnarray}
in the momentum space. Here $p_\pm=(\omega\mp k)/2$.
 Using (\ref{Dinp}) together with
(\ref{o13}), we obtain the null
 Green's function as
\begin{eqnarray}\label{d11}
&&\bar{\cal D}_{i~j-i}(p_+,p_-)=0~~~(j=i+1,\cdots,n).\nonumber
\end{eqnarray}
Also, substituting \eqref{o13} into (\ref{Dinp}) leads to the
CFT-retarded Green's function in the momentum space
\begin{eqnarray}
&&\hspace*{-1em}\bar{\cal D}_{1n}=\bar{\cal
D}_{2~n-1}=\cdots=\bar{\cal
D}_{n1}=\nonumber\\
&&\hspace*{-1em}\Big(2(\triangle-1)\Big)^n\frac{(2\pi
T_L)^{\triangle-1}(2\pi T_R)^{\triangle-1}}{\Gamma(\triangle)^2}
\sinh\left[\frac{p_+}{2T_L}
+\frac{p_-}{2T_R}\right]\nonumber\\
&&\times\Big| \Gamma\left(\frac{\triangle}{2}+i\frac{p_+}{2\pi
T_L}\right)\Big|^2 \Big|
\Gamma\left(\frac{\triangle}{2}+i\frac{p_-}{2\pi
T_R}\right)\Big|^2,\label{d1n}
\end{eqnarray}
where $\Gamma$ is the gamma function. The log-retarded Green's
functions are given by
\begin{eqnarray}
&&\hspace*{-1em}\bar{\cal D}_{2n}=\bar{\cal
D}_{3~n-1}=\cdots=\bar{\cal
D}_{n2}=\nonumber\\
&&\hspace*{-1em}\bar{\cal
D}_{1n}(p_+,p_-)\times\Bigg\{\frac{n}{\triangle-1}+\ln[2\pi
T_L]+\ln[2\pi
T_R]-2\psi(\triangle)\nonumber\\
&&+\frac{1}{2}\psi\left(\frac{\triangle}{2} +i\frac{p_+}{2\pi
T_L}\right)+\frac{1}{2}\psi\left(\frac{\triangle}{2}
-i\frac{p_+}{2\pi T_L}\right)\nonumber\\
&&+\frac{1}{2}\psi\left(\frac{\triangle}{2} +i\frac{p_-}{2\pi
T_R}\right) +\frac{1}{2}\psi\left(\frac{\triangle}{2}
-i\frac{p_-}{2\pi T_R}\right)\Bigg\}\label{d2n}
\end{eqnarray}
where
 $\psi(A)=\partial\ln[\Gamma(A)]/\partial A$ is the digamma
 function. In deriving $\bar{D}_{2n}$, we have used the following
 relation \cite{Moon:2012vc}:
 \begin{eqnarray}
\Big<{\cal O}_{i}(u_+,u_-){\cal O}_{j}(0)\Big>=\frac{1}{s!}
\left(\frac{\partial}{\partial\triangle}\right)^s\Big<{\cal
O}_{1}(u_+,u_-){\cal O}_{n}(0)\Big>.\nonumber
\end{eqnarray}
 We notice that
the log$^{n-1}$-Green's functions of $\bar{\cal D}_{kn}$$[=\bar{\cal
D}_{k+1~n-1}=\cdots=\bar{\cal D}_{nk}]$ with $k=3,\cdots,n$  can be
found along the same line using the above relation.

In order to derive quasinormal frequencies, we investigate the pole
structure of the commutators for the non-rotating BTZ black hole
($T_R=T_L=T_H=1/2\pi \ell,~r_+=\ell)$. It is found that for
$i,~j=1,2,\cdots,n$ and $s=i+j-n-1\ge 0$, the pole structure of the
retarded Green's functions is given by
\begin{eqnarray}\label{Dij}
&&\bar{\cal
D}_{ij}(p_+)\nonumber\\
&&\hspace*{-1.2em}\propto\sum_{m=0}^{s}{}_{s}C_m\Gamma^{(m)}
\left(h_L+i\frac{p_+}{2\pi
T_L}\right)\Gamma^{(s-m)}\left(h_L-i\frac{p_+}{2\pi
T_L}\right)\nonumber
\end{eqnarray}
with $\Gamma^{(m)}=\frac{\partial^m\Gamma}{\partial\triangle^m}$.
This is one of our main results. Following \cite{Birmingham:2001pj},
for the rank-$n$ case one can read off
 quasinormal frequencies of scalar $A_n$
which satisfies $(\nabla_B^2-m^2)^nA_n=0$
\begin{equation}
\omega^N_n = k- i4\pi T_L (N+h_L),~~~N=0,1,2,\cdots \label{10}
\end{equation}
from a $n$-fold pole of the retarded Green's function  $\bar{\cal
D}_{nn}(p_+)$. Applying the previous truncation process  for the
LCFT to $\bar{\cal D}_{ij}$ leads to
\begin{eqnarray}
&&\bar{\cal D}_{ij} \sim \left(
  \begin{array}{cc}
    0& 0 \\
    0 & \bar{\cal D}_{(n+1)/2~(n+1)/2} \\
  \end{array}
\right)~~~{\rm for~odd~}n,\\
&&\nonumber\\
&&\bar{\cal D}_{ij} \sim \left(
  \begin{array}{cc}
    0& ~~~~~0 \\
    0 & ~~~~~0\\
  \end{array}
\right)~~~{\rm for~even~}n,
\end{eqnarray}
which implies that for odd rank,  the matrix provides a simple pole
of $\omega^N_s = k- i4\pi T_L (N+h_L)$ obtained from  the finite
temperature CFT,  while for even rank,  it contains nothing.

 It is well-known that the absorption cross section
\cite{Gubser:1997cm} can be written in terms of frequency ($\omega$)
and temperatures ($T_{R/L},~T_H$) as
\begin{eqnarray}\label{abs0}
\sigma_{\rm abs}^{ij} &=&\frac{{\cal C}_n}{\omega}\bar{\cal
D}_{ij}(\omega),
\end{eqnarray}
where ${\cal C}_n$ is a normalization constant. Here $\bar{\cal
D}_{ij}(\omega)$ is obtained  by substituting $p_+=p_-=\omega/2$ for
$s$-wave ($k=0$) into  (\ref{d11}), (\ref{d2n}), and $\bar{\cal
D}_{pn}$$[=\bar{\cal D}_{p+1~n-1}=\cdots=\bar{\cal D}_{np}]$ with
$p=3,4,\cdots,n$. From the expression (\ref{abs0}), one can find the
absorption cross section
\begin{eqnarray}
&&\sigma_{\rm abs}^{1n}=\sigma_{\rm abs}^{2~n-1}=\cdots =\sigma_{\rm
abs}^{n1}\simeq\pi^2 \omega \ell^2,\nonumber\\
&&\sigma_{\rm abs}^{2n}=\sigma_{\rm abs}^{3~n-1}=\cdots=
~\sigma_{\rm abs}^{n2}\nonumber\\
&&\simeq \Big[
           n-2 + 2\gamma-2\ln2+2\ln[\omega \ell]\Big]
           \sigma_{\rm abs}^{1n}
\end{eqnarray}
for the low-temperature limit of $\omega \gg T_{R/L}$ and
$\triangle=2$.\\ Here the Euler's constant $\gamma=0.5772$, $\ell$
is the AdS$_3$ curvature radius, and the normalization constant is
fixed to be ${\cal C}_n=2^{1-n}$. On the other hand, one finds that
the other absorption cross sections of $\sigma_{\rm
abs}^{pn}[=\sigma_{\rm abs}^{p+1~n-1}=\cdots=\sigma_{\rm abs}^{np}]$
with $p=3,4,\cdots,n$ are given in terms of $\bar{\cal
D}_{pn}$$[=\bar{\cal D}_{p+1~n-1}=\cdots=\bar{\cal D}_{np}]$.

Finally, it turns out that in the low-temperature limit  and
$\Delta=2$, the general form of the absorption cross sections are
given by the power series expansion of $\ln[\omega\ell]$
\begin{equation}
\sigma_{\rm abs}^{ij}\vert_{\omega \gg T_{R/L}} =
\Big[\sum_{m=0}^{s}a_{sm}^{(n)}\ln^{m}[\omega\ell]\Big]\sigma_{\rm
abs}^{1n}, \label{abs-final}
\end{equation}
where $s=i+j-n-1\ge0$ and $a_{sm}^{(n)}$ are some constants to be
fixed (see arXiv:1211.3679v1). For $i,j=n$ and $s=n-1$,
(\ref{abs-final}) leads to the highest-order logarithmic correction
$\ln^{n-1}[\omega \ell]$ to the Klein-Gordon mode which corresponds
to the mode $A_n$ satisfying $\nabla_B^{2n} A_n=0$.

 Applying  the previous truncation
process to $\sigma_{\rm abs}^{ij}\vert_{\omega \gg T_{R/L}}$ leads
to
\begin{eqnarray}
&&\sigma^{ij}_{\rm abs} \sim \left(
  \begin{array}{cc}
    0& 0 \\
    0 & \sigma^{(n+1)/2~(n+1)/2}_{\rm abs} \\
  \end{array}
\right)~~~{\rm for~odd~}n,\\
&&\nonumber\\
&&\sigma^{ij}_{\rm abs} \sim \left(
  \begin{array}{cc}
    0& ~~~~~0 \\
    0 & ~~~~~0\\
  \end{array}
\right)~~~{\rm for~even~}n,
\end{eqnarray}
which imply that in the low-temperature and massless limits, the odd
rank case provides the absorption cross section for the Klein-Gordon
mode only, while for even rank,  it contains null states only.

In summary, we have constructed the rank-$n$ finite temperature LCFT
starting from the $n$-coupled scalar field theory $S_\Phi$
(\ref{sPhi}) in the BTZ black hole background. Our approach has
provided two important quantities of quasinormal frequencies and
absorption cross section of scalar $A_n$ which satisfies the $2n$-th
order linearized equation $(\nabla^2_B-m^2)^nA_n=0$ around the BTZ
black
 hole.  This work shows a usefulness of the
AdS-LCFT correspondence for obtaining two observables of quasinormal
modes and absorption cross section of $A_n$  without solving the
$2n$-th order linearized equation directly.

\hspace*{1em} This work was supported by the National Research
Foundation of Korea (NRF) grant funded by the Korea government
(MEST) through the Center for Quantum Spacetime (CQUeST) of Sogang
University with grant number 2005-0049409. Y. Myung  was partly
supported by the National Research Foundation of Korea (NRF) grant
funded by the Korea government (MEST) (No.2012-040499).

\end{document}